\documentclass[aps,nofootinbib,preprint]{revtex4}
\usepackage{color}

\newcommand{\omits}[1]{}

\begin{document}

%\preprint{hep-th/yymmnnn}

\title{Hidden Conformal Symmetry \\ of the Reissner-Nordstr{\o}m Black Holes}

\author{Chiang-Mei Chen} \email{cmchen@phy.ncu.edu.tw}
\affiliation{Department of Physics and Center for Mathematics and Theoretical Physics,
National Central University, Chungli 320, Taiwan}

\author{Jia-Rui Sun} \email{jrsun@phy.ncu.edu.tw}
\affiliation{Department of Physics, National Central University, Chungli 320, Taiwan}

\date{\today}

%% REVTEX4
%\maketitle

\begin{abstract}
Motivated by recent progresses in the holographic descriptions of
the Kerr and Reissner-Nordstr{\o}m (RN) black holes, we explore the
hidden conformal symmetry of nonextremal uplifted 5D RN black hole
by studying the near horizon wave equation of a massless scalar
field propagating in this background. Similar to the Kerr
black hole case, this hidden symmetry is broken by the periodicity
of the associated angle coordinate in the background geometry, but
the results somehow testify the dual CFT description of the
nonextremal RN black holes. The duality is further supported by
matching of the entropies and absorption cross sections
calculated from both CFT and gravity sides.
\end{abstract}

%% REVTEX4
%\pacs{}

\maketitle
%%%%%%%%%%%%%%%%%%%%%%%%%%%%%%%%%%%%%%%%%%%%%%%%%%%%%%%%%%%%%%%%%%%%%%
\tableofcontents

%%%%%%%%%%%%%%%%%%%%%%%%%%%%%%%%%%%%%%%%%%%%%%%%%%%%%%%%%%%%%%%%%%%%%%
\section{Introduction}
%%%%%%%%%%%%%%%%%%%%%%%%%%%%%%%%%%%%%%%%%%%%%%%%%%%%%%%%%%%%%%%%%%%%%%
In the past two years, the investigations on the holographic
dual descriptions \cite{'tHooft:1993gx, Susskind:1994vu,
Maldacena:1997re, Gubser:1998bc, Witten:1998qj} for the black holes
have archived substantial success, in particular for the
Kerr~\cite{Guica:2008mu, Dias:2009ex, Matsuo:2009sj,
Bredberg:2009pv, Amsel:2009pu, Hartman:2009nz, Castro:2009jf,
Cvetic:2009jn} and the Reissner-Nordstr{\o}m (RN) black
holes~\cite{Hartman:2008pb, Garousi:2009zx, Chen:2009ht,
Chen:2010bs}, as well as the other
generalizations~\cite{Hotta:2008xt, Lu:2008jk, Azeyanagi:2008kb,
Chow:2008dp, Azeyanagi:2008dk, Nakayama:2008kg, Isono:2008kx,
Peng:2009ty, Chen:2009xja, Loran:2009cr, Ghezelbash:2009gf,
Lu:2009gj, Amsel:2009ev, Compere:2009dp, Krishnan:2009tj,
Hotta:2009bm, Astefanesei:2009sh, Wen:2009qc, Azeyanagi:2009wf,
Wu:2009di, Peng:2009wx}. The major progresses are made
essentially on the extremal and near extremal limits in which the
black hole near horizon geometries consist a certain AdS structure
and the central charges of dual CFT can be obtained by analyzing the
asymptotic symmetry following the approach in~\cite{Brown:1986nw} or
by calculating the boundary stress tensor of the 2D
effective action as explicitly discussed in~\cite{Hartman:2008dq,
Alishahiha:2008tv, Castro:2008ms, Balasubramanian:2009bg,
Castro:2010vi}. The supported evidences include the matching
of the CFT and black hole entropies, and the agreement between the
absorption cross section of a scalar field with the two point
function of its dual operator.

It is naturally to expect that the holographic dual
description should be legitimate for the general nonextremal black
holes. However, there are no obvious hints for the AdS structure in
the near horizon geometry of nonextremal black holes. The conformal
symmetry somehow has been hidden and therefore the study of the
corresponding dual CFT seems to be obscure. In the recent
paper~\cite{Castro:2010fd} an encouraging result has been shown that
a massless scalar field is indeed able to probe the hidden
conformal symmetry in the nonextremal Kerr black hole background, 
see also~\cite{Krishnan:2010pv}. The work~\cite{Castro:2010fd} 
is motivated by the fact that the near horizon
wave equation for a massless scalar field comprehend a conformal
symmetry inherited from the back hole background geometry. In this
approach, the dual CFT temperatures of left and right sectors can be
identified. Moreover, the essential information of the operator dual
to the scalar field can also be read out and the absorption cross 
section of the massless scalar field agrees with the two
point function of its dual operator. However, in the background
geometry, the conformal symmetry is broken to $U(1)_L \times U(1)_R$
due to the periodicity of the corresponding angle coordinate.

In this paper, we investigate the hidden conformal symmetry
of RN black holes following the approach in~\cite{Castro:2010fd}.
The RN black holes can have a holographic dual CFT$_2$ description
by uplifting to the 5D spacetime. The central charges generically
depend on the radius of the extra dimension $\ell$ as: $c_L = c_R =
6 q^3/\ell$. By analyzing the near horizon Klein-Gorden equation of
a massless scalar field in the low frequency, $\omega \ll 1/m$ and
low momentum $k/\ell \ll 1/m$ limit, we found that the wave equation
can be derived from the Casimir operator of CFT$_2$. Consequently,
the finite temperatures of left and right sectors are $T_L = \ell (2
m^2 - q^2) / 2 \pi q^3$ and $T_R = \ell m \sqrt{m^2 - q^2}/\pi q^3$
respectively, and the CFT entropy precisely reproduces the black
hole entropy
$$
S_\mathrm{CFT} = \frac{\pi^2}{3} \, \frac{6 q^3}{\ell} \left[ \frac{\ell (2 m^2 - q^2)}{2 \pi q^3} + \frac{\ell m \sqrt{m^2 - q^2}}{\pi q^3} \right] = \pi r_+^2 = \frac{\mathrm{Area}}4.
$$
We also check the agreement of the absorption cross section with the two point function of corresponding dual operator.

The outline of this paper is as follows. We review the uplifted RN black holes
and discuss the associated properties in Section II. In Section III,
we derive the near horizon wave equation for a massless scalar field
and explore the hidden conformal symmetry in the equation. The CFT temperatures
of left and right sectors are identified. We further, in Section IV, verify the absorption
cross section of the massless scalar field and the two point function of its
dual operator. Finally summarize our results in Section V.

%%%%%%%%%%%%%%%%%%%%%%%%%%%%%%%%%%%%%%%%%%%%%%%%%%%%%%%%%%%%%%%%%%%%%%
\section{Uplifted 5D RN Black Hole}
%%%%%%%%%%%%%%%%%%%%%%%%%%%%%%%%%%%%%%%%%%%%%%%%%%%%%%%%%%%%%%%%%%%%%%
We first review the properties of 5D RN black hole uplifted from its
4D counterpart. Note that the electrically charged RN black hole in
the 4D Einstein-Maxwell theory
\begin{equation}
I_4 = \frac1{16\pi G_4} \int d^4x \sqrt{-g} \left( R - \frac14 F_{[2]}^2 \right),
\end{equation}
of metric and gauge potential ($m, \; q$ are mass and charge parameters)
\begin{eqnarray}\label{RN}
ds_4^2 &=& - f(r) dt^2 + \frac{dr^2}{f(r)} + r^2 d\Omega_2^2,
\nonumber\\
A_{[1]} &=& - \frac{2 q}{r} dt, \qquad f(r) = 1 - \frac{2m}{r} + \frac{q^2}{r^2},
\end{eqnarray}
can be consistently uplifted to 5D Einstein-Maxwell theory
\begin{equation}
I_5 = \frac1{16\pi G_5} \int d^5x \sqrt{- \hat g} \left( \hat R -
\frac1{12} \hat F_{[3]}^2 \right),
\end{equation}
via an inverse Kaluza-Klein (KK) reduction~\cite{Chen:2010bs}
\begin{equation}
ds_5^2 = \left( \ell d\chi + \frac12 A_\mu dx^\mu \right)^2 +
ds_4^2, \qquad \hat A_{[2]} = \frac{\sqrt3}2 A_{[1]} \wedge \ell
d\chi.
\end{equation}
The explicit form of uplifted RN solution is
\begin{eqnarray}\label{RN5metric}
ds^2 &=& - f(r) dt^2 + \frac{dr^2}{f(r)} + r^2 d\Omega_2^2 +
\left(\ell d\chi - \frac{q}{r} dt \right)^2,
\nonumber\\
\hat A_{[2]} &=& - \sqrt3 \, \frac{q}{r} \, dt \wedge \ell d\chi.
\end{eqnarray}
The radius of extra dimension of a circle, the parameter $\ell$, is
explicitly given in the solution to ensure the usual period of extra
angle coordinate $\chi \sim \chi + 2 \pi$. The gravitational
constants in two different dimensions are related by $G_5 = 2 \pi
\ell G_4$. Hereafter we will assume $G_4 = 1$. A notable feature of
the uplifted RN black hole is the emergence of the ergosphere at $-
g_{tt} = 1 - 2 m /r = 0$ which is an essential feature for superradiance study~\cite{Chen:2010bs}.
The corresponding black hole thermodynamic
quantities, such as the Hawking temperature and the Bekenstein-Hawking
entropy, are
\begin{eqnarray}
T_H &=& \frac1{4\pi} \frac{r_+ - r_-}{r_+^2},
\nonumber\\
S_{BH} &=& \frac{A_5}{4 G_5} = \frac{A_4}{4}= \pi r_+^2,
\end{eqnarray}
where $r_\pm = m \pm \sqrt{m^2 - q^2}$ are the black hole outer and inner horizon
radius, respectively.

The dual CFT descriptions of the near extremal RN black hole have been studied both
in the (warped) AdS$_3$/CFT$_2$~\cite{Hartman:2008pb, Garousi:2009zx, Chen:2010bs}
and AdS$_2$/CFT$_1$ pictures~\cite{Chen:2009ht}. The central charge of the CFT can be
computed by analyzing the asymptotic symmetry of the near horizon
geometry in the extremal limit. It is shown that, unlike the Kerr
black hole case, the central charges for uplifted RN black holes
depend on the choice of parameter $\ell$. In particular, two
specific choices have been discussed: $c_L = c_R = 6 q^3$ for $\ell
= 1$ in~\cite{Hartman:2008pb, Garousi:2009zx} and $c_L = c_R = 6
q^2$ for $\ell = q$ in~\cite{Chen:2009ht, Chen:2010bs}. Accordingly,
the general expression for central charges can be written as
\begin{equation}
c_L = c_R = \frac{6 q^3}{\ell}.
\end{equation}

%%%%%%%%%%%%%%%%%%%%%%%%%%%%%%%%%%%%%%%%%%%%%%%%%%%%%%%%%%%%%%%%%%%%%%
\section{Scalar Field Equation}
%%%%%%%%%%%%%%%%%%%%%%%%%%%%%%%%%%%%%%%%%%%%%%%%%%%%%%%%%%%%%%%%%%%%%%
Consider a bulk massless scalar field $\Phi$ propagating in the
background of (\ref{RN5metric}), the Klein-Gordon (KG) equation
\begin{equation} \label{KG}
\nabla_\alpha \nabla^\alpha \Phi = 0,
\end{equation}
can be simplified by assuming the following form of the scalar field
\begin{equation} \label{APhi5}
\Phi(t, r, \theta, \phi, \chi) = \mathrm{e}^{-i \omega t + i n \phi + i k \chi} S(\theta) R(r),
\end{equation}
and reduces to two decoupled equations by separation of variables:
\begin{eqnarray}
\partial_r (\Delta \partial_r R) + \left[ \frac{(\omega r - k q/\ell)^2 r^2}{\Delta} - k^2/\ell^2 r^2 - \lambda_l \right] R &=& 0,
\label{EqRr} \\
\frac1{\sin\theta} \partial_\theta (\sin\theta \partial_\theta S_l) + \left( \lambda_l - \frac{n^2}{\sin^2\theta} \right) S_l &=& 0,
\label{EqS}
\end{eqnarray}
where $\Delta = r^2 f = (r - r_+) (r - r_-)$. The angular equation simply implies that the separation constant should take the standard value $\lambda_l = l (l + 1)$ for integer $l$ and the solutions for $S_l$ are just the standard spherical harmonic functions.

It is worth to note that the probe of a neutral massless scalar field in 5D uplifted RN black holes is equivalent to a probe in 4D RN black holes by a charged massive scalar field with mass $\bar\mu$ and charge $e$ given by~\cite{Chen:2010bs}
\begin{equation}
e = \frac12 \frac{k}{\ell}, \qquad \bar\mu^2 = \frac{k^2}{\ell^2}.
\end{equation}
Therefore, the 5D momentum mode $k$ generates both the charge and mass of the 4D scalar. The factor half is due to the $k$-momentum coupling only with the KK vector portion, namely half of the original 4D RN background charge.

Furthermore, the radial equation can be reformulated in the following form
\begin{eqnarray}\label{scalarEQ}
&&\partial_r (\Delta \partial_r R) + \Biggl[ \frac{\left[ (2 m r_+ -
q^2) \omega - (q r_+/\ell) k \right]^2}{(r - r_+)(r_+ - r_-)} -
\frac{\left[ (2 m r_- - q^2) \omega - (q r_-/\ell) k \right]^2}{(r -
r_-)(r_+ - r_-)}
\nonumber\\
&&+ (\omega^2 - k^2/\ell^2) (r^2 - q^2) + 2 \omega (\omega m - k
q/\ell) (r + 2m) \Biggr] R = l (l+1) R.
\end{eqnarray}
We will see later that if the potential term in second line of
Eq.(\ref{scalarEQ}) vanishes, then the remaining equation can be exactly obtained by the Casimir
operator of the $SL(2,R)_L \times SL(2,R)_R$ Lie algebra. To
accomplish this, we should require: (a) small frequency $\omega$ and momentum $k/\ell$
limit, namely $m \omega \ll 1, \; m k/\ell \ll 1$ (automatically
implies $q \omega \ll 1, \; q k/\ell \ll 1$), and (b) near horizon
limit of $r \omega \ll 1$ and $r k/\ell \ll 1$. Consequently, the
near-horizon radial wave equation reduces to
\begin{equation} \label{NHrEQ}
\partial_r (\Delta \partial_r R) + \Biggl[ \frac{\left[ (2 m r_+ - q^2) \omega - (q r_+/\ell) k \right]^2}{(r - r_+)(r_+ - r_-)} - \frac{\left[ (2 m r_- - q^2) \omega - (q r_-/\ell) k \right]^2}{(r - r_-)(r_+ - r_-)} \Biggr] R = l (l+1) R.
\end{equation}

The two sets of symmetry generators of the AdS$_3$ space with radius
$L$, in the Poincar\'e coordinates: ($w^\pm, y$),
\begin{equation}
ds_3^2 = \frac{L^2}{y^2} ( dy^2 + dw^+ dw^-),
\end{equation}
are
\begin{eqnarray}
H_1 &=& i \partial_+,
\nonumber\\
H_0 &=& i \left( w^+ \, \partial_+ + \frac12 y \, \partial_y \right),
\nonumber\\
H_{-1} &=& i \left( (w^+)^2 \, \partial_+ + w^+ y \, \partial_y - y^2 \, \partial_- \right),
\end{eqnarray}
and
\begin{eqnarray}
\bar H_1 &=& i \partial_-,
\nonumber\\
\bar H_0 &=& i \left( w^- \, \partial_- + \frac12 y \, \partial_y \right),
\nonumber\\
\bar H_{-1} &=& i \left( (w^-)^2 \, \partial_- + w^- y \, \partial_y - y^2 \, \partial_+ \right),
\end{eqnarray}
assembling two copies of the $SL(2,R)$ Lie algebra
\begin{equation}
\left[ H_0, H_{\pm 1} \right] = \mp i H_{\pm 1}, \qquad \left[ H_{-1}, H_1 \right] = - 2 i H_0.
\end{equation}
Thus the corresponding Casimir operator is
\begin{equation}
\mathcal{H}^2 = \bar\mathcal{H}^2 = - H_0^2 + \frac12 \left( H_1 H_{-1} + H_{-1} H_1 \right) = \frac14 \left( y^2 \, \partial_y^2 - y \, \partial_y \right) + y^2 \, \partial_+ \partial_-.
\end{equation}

Converting the Poincar\'e coordinates ($w^\pm, y$) to the
coordinates ($t, r, \chi$) of uplifted RN black hole by the
following transformations
\begin{eqnarray}
w^+ &=& \sqrt{\frac{r - r_+}{r - r_-}} \, \exp(2 \pi T_R \chi + 2 n_+ t),
\nonumber\\
w^- &=& \sqrt{\frac{r - r_+}{r - r_-}} \, \exp(2 \pi T_L \chi + 2 n_- t),
\nonumber\\
y &=& \sqrt{\frac{r_+ - r_-}{r - r_-}} \, \exp[\pi (T_R + T_L) \chi + (n_+ + n_-) t],
\end{eqnarray}
where
\begin{equation}
T_R = \frac{(r_+ - r_-) m \ell}{2 \pi q^3}, \qquad T_L = \frac{(r_+ + r_-) m \ell}{2 \pi q^3} - \frac{\ell}{2 \pi q}, \qquad n_\pm = - \frac{r_+ \mp r_-}{4 q^2},
\end{equation}
we can directly calculate all the $SL(2,R)$ generators in terms of black hole coordinates
\begin{eqnarray}
H_1 &=& i \mathrm{e}^{-(2 \pi T_R \chi + 2 n_+ t)} \left( \sqrt{\Delta} \, \partial_r + \frac{m^2}{\pi q^2 T_R} \frac{r - q^2/m}{\sqrt{\Delta}} \, \partial_\chi + \frac{2 T_L}{T_R} \frac{m r - \frac{m^2 q^2}{2m^2 - q^2}}{\sqrt{\Delta}} \, \partial_t \right),
\nonumber\\
H_0 &=& i \frac{m^2}{\pi q^2 T_R} \, \partial_\chi + i \frac{2 m T_L}{T_R} \, \partial_t,
\nonumber\\
H_{-1} &=& i \mathrm{e}^{2 \pi T_R \chi + 2 n_+ t} \left(- \sqrt{\Delta} \, \partial_r + \frac{m^2}{\pi q^2 T_R} \frac{r - q^2/m}{\sqrt{\Delta}} \, \partial_\chi + \frac{2 T_L}{T_R} \frac{m r - \frac{m^2 q^2}{2m^2 - q^2}}{\sqrt{\Delta}} \, \partial_t \right),
\end{eqnarray}
and
\begin{eqnarray}
\bar H_1 &=& - i \mathrm{e}^{-(2 \pi T_L \chi + 2 n_- t)} \left( \sqrt{\Delta} \, \partial_r - \frac{2m^2-q^2}{2 \pi q^2 T_L} \frac{r}{\sqrt{\Delta}} \, \partial_\chi - \frac{2 m r - q^2}{\sqrt{\Delta}} \, \partial_t \right),
\nonumber\\
\bar H_0 &=& i \frac{2m^2 - q^2}{2 \pi q^2 T_L} \, \partial_\chi + i 2m \, \partial_t,
\nonumber\\
\bar H_{-1} &=&  - i \mathrm{e}^{2 \pi T_L \chi + 2 n_- t} \left( - \sqrt{\Delta} \, \partial_r - \frac{2m^2-q^2}{2 \pi q^2 T_L} \frac{r}{\sqrt{\Delta}} \, \partial_\chi - \frac{2 m r - q^2}{\sqrt{\Delta}} \, \partial_t \right).
\end{eqnarray}
Finally the Casimir operator becomes
\begin{equation}
\mathcal{H}^2 = \partial_r \Delta \partial_r - \frac{\left[ (2 m r_+
- q^2) \partial_t + (q r_+ /\ell) \partial_\chi \right]^2}{(r -
r_+)(r_+ - r_-)} + \frac{\left[ (2 m r_- - q^2) \partial_t + (q r_-/
\ell) \partial_\chi \right]^2}{(r - r_-)(r_+ - r_-)}.
\end{equation}
Therefore, the near-horizon wave equation Eq.(\ref{NHrEQ}) can be
formulated as
\begin{equation}
\bar \mathcal{H}^2 \Phi = \mathcal{H}^2 \Phi = l (l+1) \Phi,
\end{equation}
and the conformal weights of dual operator of the massless field $\Phi$ should be
\begin{equation}
(h_L, h_R) = (l + 1, l + 1).
\end{equation}
The microscopic entropy of the dual CFT can be computed by the Cardy
formula which matches with the black hole Bekenstein-Hawking entropy
\begin{equation}
S_\mathrm{CFT} = \frac{\pi^2}3 (c_L T_L + c_R T_R) = \pi (2 m r_+ - q^2) = \pi r_+^2 = S_\mathrm{BH}.
\end{equation}

The background geometry of the nonextremal uplifted RN black holes, however, does not consist the full $SL(2,R)_L \times SL(2,R)_R$ symmetry. Actually, this conformal symmetry is broken down to $U(1)_L \times U(1)_R$ by the periodicity of the angle coordinate $\chi$. The $SL(2,R)$ generators are not periodic under the angular identification $\chi \sim \chi + 2 \pi$, but are transformed to
\begin{equation}
w^+ \sim \mathrm{e}^{4\pi^2 T_R} \, w^+, \qquad w^- \sim \mathrm{e}^{4\pi^2 T_L} \, w^-, \qquad y \sim \mathrm{e}^{2\pi^2(T_L + T_R)} \, y,
\end{equation}
which just corresponding to operation of the $U(1)_L \times U(1)_R$ group element in $SL(2,R)_L \times SL(2,R)_R$
\begin{equation}
\mathrm{e}^{-i 4\pi^2 (T_L \bar H_0 + T_R H_0)}.
\end{equation}

%%%%%%%%%%%%%%%%%%%%%%%%%%%%%%%%%%%%%%%%%%%%%%%%%%%%%%%%%%%%%%%%%%%%%%
\section{Scattering}
%%%%%%%%%%%%%%%%%%%%%%%%%%%%%%%%%%%%%%%%%%%%%%%%%%%%%%%%%%%%%%%%%%%%%%
The scattering process of a massive scalar field in the uplifted RN black
hole background (\ref{RN5metric}) has been discussed in
\cite{Chen:2010bs} in the near extremal limit, and the absorption
cross section calculated from the geometric side matches with the
microscopic greybody factor calculated from the dual CFT side. In
this section, we will extend the computation into nonextremal limit of the uplifted RN
black hole. The near horizon region solutions, for $r \ll 1/\omega, \, r \ll \ell/k$, of Eq.(\ref{NHrEQ}) include both ingoing and outgoing modes
\begin{eqnarray}
R^\mathrm{(in)} &=& \left( \frac{r-r_+}{r-r_-} \right)^{- i \frac{r_+ (2 m \omega - q \mu \omega - q \tilde k)}{r_+ - r_-}} (r - r_-)^{-l-1}
\\
&& F\left( 1 \!+\! l \!-\! \frac{2(2 m^2 \omega \!-\! q^2 \omega \!-\! m q \tilde k)}{r_+ - r_-}, 1 \!+\! l \!-\! i (2 m \omega \!-\! q \tilde k); 1 \!-\! i \frac{2 r_+ (2 m \omega \!-\! q \mu \omega \!-\! q \tilde k)}{r_+ - r_-}; \frac{r \!-\! r_+}{r \!-\! r_-} \right),
\nonumber\\
R^\mathrm{(out)} &=& \left( \frac{r-r_+}{r-r_-} \right)^{i \frac{r_+ (2 m \omega - q \mu \omega - q \tilde k)}{r_+ - r_-}} (r - r_-)^{-l-1}
\\
&& F\left( 1 \!+\! l \!+\! \frac{2(2 m^2 \omega \!-\! q^2 \omega \!-\! m q \tilde k)}{r_+ - r_-}, 1 \!+\! l \!+\! i (2 m \omega \!-\! q \tilde k); 1 \!+\! i \frac{2 r_+ (2 m \omega \!-\! q \mu \omega \!-\! q \tilde k)}{r_+ - r_-}; \frac{r \!-\! r_+}{r \!-\! r_-} \right), \nonumber
\end{eqnarray}
where $\tilde k = k/\ell$ and $\mu = q/r_+$ is the chemical
potential. At the outer boundary of the matching region $r \gg m$
(but still $r \ll 1/\omega, \, r \ll 1/\tilde k$)
\begin{equation}
R^\mathrm{(in)}(r \gg m) \sim A r^l + B r^{-l - 1},
\end{equation}
with
\begin{eqnarray}
A &=& \frac{\Gamma\left( 1 - i \frac{2 r_+ (2 m \omega - q \mu \omega - q \tilde k)}{r_+ - r_-} \right) \Gamma(1 + 2 l)}{\Gamma\left(1 + l - i (2 m \omega - q \tilde k) \right) \Gamma\left( 1 + l - i \frac{2(2 m^2 \omega - q^2 \omega - m q \tilde k)}{r_+ - r_-} \right)},
\\
B &=& \frac{\Gamma\left( 1 - i \frac{2 r_+ (2 m \omega - q \mu \omega - q \tilde k)}{r_+ - r_-} \right) \Gamma(- 1 - 2 l)}{\Gamma\left( - l - i (2 m \omega - q \tilde k) \right) \Gamma\left( - l - i \frac{2(2 m^2 \omega - q^2 \omega - m q \tilde k)}{r_+ - r_-} \right)},
\end{eqnarray}
up to an overall constant independent of $\omega$ and $k$. Formally
we should also find the asymptotic region solutions, for $r \gg m$,
and then match two sets of solutions at the overlap region, i.e. $m
\ll r \ll (1/\omega, 1/\tilde k)$, in order to get the relations
among the integration constants~\cite{Chen:2010bs, Bredberg:2009pv}.
However, the essential properties of the absorption cross section
is indeed captured by the coefficient $A$ such as
\begin{eqnarray}
P_\mathrm{abs} &\sim& |A|^{-2}
\nonumber\\
&\sim&
%\frac{r_+ - r_-}{2 \pi r_+ (2 m \omega - q \mu \omega - q \tilde k)}
\sinh\left( \frac{2 \pi r_+ (2 m \omega - q \mu \omega - q \tilde k)}{r_+ - r_-} \right)
\nonumber\\
&& \left| \Gamma\left(1 + l - i (2 m \omega - q \tilde k) \right) \right|^2 \; \left| \Gamma\left( 1 + l - i \frac{2(2 m^2 \omega - q^2 \omega - m q \tilde k)}{r_+ - r_-} \right) \right|^2.
\end{eqnarray}

To see explicitly that $P_\mathrm{abs}$ matches with the microscopic
greybody factor of the dual CFT, we need to identify the related parameters.
From the first law of black hole thermodynamics
\begin{equation}
T_H \delta S_{BH} = \delta m - \mu \delta q.
\end{equation}
one can compute the conjugate charges as
\begin{equation}
\delta S_{BH} = \delta S_{CFT} = \frac{\delta E_L}{T_L} + \frac{\delta E_R}{T_R},
\end{equation}
and the solution is
\begin{eqnarray}
\delta E_L &=& \frac{(2m^2 - q^2) (2 m \ell \delta m - q \delta q)}{q^3},
\nonumber\\
\delta E_R &=& \frac{2 (2m^2 - q^2) m \ell \delta m - 2 m^2 q \delta q}{q^3}.
\end{eqnarray}
Therefore, the frequencies of left and right sectors can be read out by identifying $\omega = \delta m, \, k = \delta q$, namely
\begin{eqnarray}
\omega_L &=& \delta E_L = \frac{(2m^2 - q^2) (2 m \ell \omega - q k)}{q^3},
\nonumber\\
\omega_R &=& \delta E_R = \frac{2 (2m^2 - q^2) m \ell \omega - 2 m^2 q k}{q^3}.
\end{eqnarray}
Finally, the absorption cross section can be expressed as
\begin{equation}\label{CFTp}
P_\mathrm{abs} \sim T_L^{2h_L - 1} T_R^{2h_R - 1} \sinh\left( \frac{\omega_L}{2 T_L} + \frac{\omega_R}{2 T_R} \right) \left| \Gamma\left( h_L + i \frac{\omega_L}{2 \pi T_L} \right) \right|^2 \, \left| \Gamma\left( h_R + i \frac{\omega_R}{2 \pi T_R} \right) \right|^2,
\end{equation}
which is the finite temperature absorption cross section for
a 2D CFT.

%%%%%%%%%%%%%%%%%%%%%%%%%%%%%%%%%%%%%%%%%%%%%%%%%%%%%%%%%%%%%%%%%%%%%%
\section{Conclusion}
%%%%%%%%%%%%%%%%%%%%%%%%%%%%%%%%%%%%%%%%%%%%%%%%%%%%%%%%%%%%%%%%%%%%%%
The $\mathrm{AdS}_3/\mathrm{CFT}_2$ description for the RN black
holes can be discussed by considering their uplifted counterparts.
The central charges of the dual CFT can be computed by analyzing the
near horizon geometry in the extremal limit. It turns out that the
central charges will depend on the embedding, namely depend on
the radius of the extra dimension. The holographic dual
descriptions of RN black holes have been investigated for the
extremal~\cite{Hartman:2008pb, Garousi:2009zx, Chen:2009ht} and near
extremal~\cite{Chen:2010bs} limits. For the general nonextremal
case, however, the geometry does not consist obvious AdS structure.
Actually, the conformal symmetry is broken by the periodicity of the
corresponding angle coordinate. Nevertheless, we found that
a massless scalar field can probe this hidden conformal symmetry in
the near horizon region, and by a suitable coordinates
identification, the Casimir operator of $SL(2,R)_L \times SL(2,R)_R$
reproduces the Klein-Gorden equation of the massless scalar field in
certain limits. Therefore we can derive the general expressions for
the dual CFT temperatures of left and right sectors and the Cardy
formula for the CFT entropy precisely agree with the black hole
entropy. Moreover, we also can identify the conformal weights and
frequencies of the operator dual to the massless scalar field. The
further support for the holographic duality is the evidence that the
absorption cross section computed from the gravity side agrees with
the two point function of operators in the CFT.

%%%%%%%%%%%%%%%%%%%%%%%%%%%%%%%%%%%%%%%%%%%%%%%%%%%%%%%%%%%%%%%%%%%%%%
\section*{Acknowledgement}
%%%%%%%%%%%%%%%%%%%%%%%%%%%%%%%%%%%%%%%%%%%%%%%%%%%%%%%%%%%%%%%%%%%%%%
This work was supported by the National Science Council of the R.O.C. under 
the grant NSC 96-2112-M-008-006-MY3 and in part by the National Center of 
Theoretical Sciences (NCTS).

%%%%%%%%%%%%%%%%%%%%%%%%%%%%%%%%%%%%%%%%%%%%%%%%%%%%%%%%%%%%%%%%%%%%%%


\begin{references}
%%%%%%%%%%%%%%%%%%%%%%%%%%%%%%%%%%%%%%%%%%%%%%%%%%%%%%%%%%%%%%%%%%%%%%


%%%%  Holographic  %%%%%%%%%%%%%%%%%%%%%%%%%%%%%%%%%%%%%%%%%%%%%%%%%%%

%\cite{'tHooft:1993gx}
\bibitem{'tHooft:1993gx}
  G.~'t Hooft,
  ``Dimensional reduction in quantum gravity,''
  arXiv:gr-qc/9310026.
  %%CITATION = GR-QC/9310026;%%

%\cite{Susskind:1994vu}
\bibitem{Susskind:1994vu}
  L.~Susskind,
  ``The world as a hologram,''
  J.\ Math.\ Phys.\  {\bf 36}, 6377 (1995)
  [arXiv:hep-th/9409089].
  %%CITATION = JMAPA,36,6377;%%

%\cite{Maldacena:1997re}
\bibitem{Maldacena:1997re}
  J.~M.~Maldacena,
  ``The large N limit of superconformal field theories and supergravity,''
  Adv.\ Theor.\ Math.\ Phys.\  {\bf 2}, 231 (1998)
  [Int.\ J.\ Theor.\ Phys.\  {\bf 38}, 1113 (1999)]
  [arXiv:hep-th/9711200].
  %%CITATION = IJTPB,38,1113;%%

%\cite{Gubser:1998bc}
\bibitem{Gubser:1998bc}
  S.~S.~Gubser, I.~R.~Klebanov and A.~M.~Polyakov,
  ``Gauge theory correlators from non-critical string theory,''
  Phys.\ Lett.\  B {\bf 428}, 105 (1998)
  [arXiv:hep-th/9802109].
  %%CITATION = PHLTA,B428,105;%%

%\cite{Witten:1998qj}
\bibitem{Witten:1998qj}
  E.~Witten,
  ``Anti-de Sitter space and holography,''
  Adv.\ Theor.\ Math.\ Phys.\  {\bf 2}, 253 (1998)
  [arXiv:hep-th/9802150].
  %%CITATION = 00203,2,253;%%


%%%  Kerr/CFT  %%%%%%%%%%%%%%%%%%%%%%%%%%%%%%%%%%%%%%%%%%%%%%%%%%%%%%%

%\cite{Guica:2008mu}
\bibitem{Guica:2008mu}
  M.~Guica, T.~Hartman, W.~Song and A.~Strominger,
  ``The Kerr/CFT correspondence,''
  Phys.\ Rev.\  D {\bf 80}, 124008 (2009)
  [arXiv:0809.4266 [hep-th]].
  %%CITATION = PHRVA,D80,124008;%%

%\cite{Dias:2009ex}
\bibitem{Dias:2009ex}
  O.~J.~C.~Dias, H.~S.~Reall and J.~E.~Santos,
  ``Kerr-CFT and gravitational perturbations,''
  JHEP {\bf 0908}, 101 (2009)
  [arXiv:0906.2380 [hep-th]].
  %%CITATION = JHEPA,0908,101;%%

%\cite{Matsuo:2009sj}
\bibitem{Matsuo:2009sj}
  Y.~Matsuo, T.~Tsukioka and C.~M.~Yoo,
  ``Another Realization of Kerr/CFT Correspondence,''
  Nucl.\ Phys.\  B {\bf 825}, 231 (2010)
  [arXiv:0907.0303 [hep-th]].
  %%CITATION = NUPHA,B825,231;%%

%\cite{Bredberg:2009pv}
\bibitem{Bredberg:2009pv}
  I.~Bredberg, T.~Hartman, W.~Song and A.~Strominger,
  ``Black Hole Superradiance From Kerr/CFT,''
  JHEP {\bf 1004}, 019 (2010)
  [arXiv:0907.3477 [hep-th]].
  %%CITATION = JHEPA,1004,019;%%

%\cite{Amsel:2009pu}
\bibitem{Amsel:2009pu}
  A.~J.~Amsel, D.~Marolf and M.~M.~Roberts,
  ``On the Stress Tensor of Kerr/CFT,''
  JHEP {\bf 0910}, 021 (2009)
  [arXiv:0907.5023 [hep-th]].
  %%CITATION = JHEPA,0910,021;%%

%\cite{Hartman:2009nz}
\bibitem{Hartman:2009nz}
  T.~Hartman, W.~Song and A.~Strominger,
  ``Holographic Derivation of Kerr-Newman Scattering Amplitudes for General Charge and Spin,''
  JHEP {\bf 1003}, 118 (2010)
  [arXiv:0908.3909 [hep-th]].
  %%CITATION = JHEPA,1003,118;%%

%\cite{Castro:2009jf}
\bibitem{Castro:2009jf}
  A.~Castro and F.~Larsen,
  ``Near Extremal Kerr Entropy from $AdS_2$ Quantum Gravity,''
  JHEP {\bf 0912}, 037 (2009)
  [arXiv:0908.1121 [hep-th]].
  %%CITATION = JHEPA,0912,037;%%

%\cite{Cvetic:2009jn}
\bibitem{Cvetic:2009jn}
  M.~Cvetic and F.~Larsen,
  ``Greybody Factors and Charges in Kerr/CFT,''
  JHEP {\bf 0909}, 088 (2009)
  [arXiv:0908.1136 [hep-th]].
  %%CITATION = JHEPA,0909,088;%%


%%%% RN/CFT  %%%%%%%%%%%%%%%%%%%%%%%%%%%%%%%%%%%%%%%%%%%%%%%%%%%%%%%%%

%\cite{Hartman:2008pb}
\bibitem{Hartman:2008pb}
  T.~Hartman, K.~Murata, T.~Nishioka and A.~Strominger,
  ``CFT duals for extreme black holes,''
  JHEP {\bf 0904}, 019 (2009)
  [arXiv:0811.4393 [hep-th]].
  %%CITATION = JHEPA,0904,019;%%

%\cite{Garousi:2009zx}
\bibitem{Garousi:2009zx}
  M.~R.~Garousi and A.~Ghodsi,
  ``The RN/CFT Correspondence,''
  Phys.\ Lett.\  B {\bf 687}, 79 (2010)
  [arXiv:0902.4387 [hep-th]].
  %%CITATION = PHLTA,B687,79;%%

%\cite{Chen:2009ht}
\bibitem{Chen:2009ht}
  C.~M.~Chen, J.~R.~Sun and S.~J.~Zou,
  ``The RN/CFT Correspondence Revisited,''
  JHEP {\bf 1001}, 057 (2010)
  [arXiv:0910.2076 [hep-th]].
  %%CITATION = JHEPA,1001,057;%%

%\cite{Chen:2010bs}
\bibitem{Chen:2010bs}
  C.~M.~Chen, Y.~M.~Huang and S.~J.~Zou,
  ``Holographic Duals of Near-extremal Reissner-Nordstrom Black Holes,''
  JHEP {\bf 1003}, 123 (2010)
  [arXiv:1001.2833 [hep-th]].
  %%CITATION = JHEPA,1003,123;%%


%%%% More Generalization  %%%%%%%%%%%%%%%%%%%%%%%%%%%%%%%%%%%%%%%%%%%%

%\cite{Hotta:2008xt}
\bibitem{Hotta:2008xt}
  K.~Hotta, Y.~Hyakutake, T.~Kubota, T.~Nishinaka and H.~Tanida,
  ``The CFT-interpolating Black Hole in Three Dimensions,''
  JHEP {\bf 0901}, 010 (2009)
  [arXiv:0811.0910 [hep-th]].
  %%CITATION = JHEPA,0901,010;%%

%\cite{Lu:2008jk}
\bibitem{Lu:2008jk}
  H.~Lu, J.~Mei and C.~N.~Pope,
  ``Kerr/CFT Correspondence in Diverse Dimensions,''
  JHEP {\bf 0904}, 054 (2009)
  [arXiv:0811.2225 [hep-th]].
  %%CITATION = JHEPA,0904,054;%%

%cite{Azeyanagi:2008kb}
\bibitem{Azeyanagi:2008kb}
  T.~Azeyanagi, N.~Ogawa and S.~Terashima,
  ``Holographic Duals of Kaluza-Klein Black Holes,''
  JHEP {\bf 0904}, 061 (2009)
  [arXiv:0811.4177 [hep-th]].
  %%CITATION = JHEPA,0904,061;%%

%\cite{Chow:2008dp}
\bibitem{Chow:2008dp}
  D.~D.~K.~Chow, M.~Cvetic, H.~Lu and C.~N.~Pope,
  ``Extremal Black Hole/CFT Correspondence in (Gauged) Supergravities,''
  Phys.\ Rev.\  D {\bf 79}, 084018 (2009)
  [arXiv:0812.2918 [hep-th]].
  %%CITATION = PHRVA,D79,084018;%%

%\cite{Azeyanagi:2008dk}
\bibitem{Azeyanagi:2008dk}
  T.~Azeyanagi, N.~Ogawa and S.~Terashima,
  ``The Kerr/CFT Correspondence and String Theory,''
  Phys.\ Rev.\  D {\bf 79}, 106009 (2009)
  [arXiv:0812.4883 [hep-th]].
  %%CITATION = PHRVA,D79,106009;%%

\bibitem{Nakayama:2008kg}
  Y.~Nakayama,
  ``Emerging AdS from extremally rotating NS5-branes,''
  Phys.\ Lett.\  B {\bf 673}, 272 (2009)
  [arXiv:0812.2234 [hep-th]].
  %%CITATION = PHLTA,B673,272;%%

%\cite{Isono:2008kx}
\bibitem{Isono:2008kx}
  H.~Isono, T.~S.~Tai and W.~Y.~Wen,
  ``Kerr/CFT correspondence and five-dimensional BMPV black holes,''
  Int.\ J.\ Mod.\ Phys.\  A {\bf 24}, 5659 (2009)
  [arXiv:0812.4440 [hep-th]].
  %%CITATION = IMPAE,A24,5659;%%

%\cite{Peng:2009ty}
\bibitem{Peng:2009ty}
  J.~J.~Peng and S.~Q.~Wu,
  ``Extremal Kerr black hole/CFT correspondence in the five dimensional G\'odel universe,''
  Phys.\ Lett.\  B {\bf 673}, 216 (2009)
  [arXiv:0901.0311 [hep-th]].
  %%CITATION = PHLTA,B673,216;%%

%\cite{Chen:2009xja}
\bibitem{Chen:2009xja}
  C.~M.~Chen and J.~E.~Wang,
  ``Holographic Duals of Black Holes in Five-dimensional Minimal Supergravity,''
  Class.\ Quant.\ Grav.\  {\bf 27}, 075004 (2010)
  [arXiv:0901.0538 [hep-th]].
  %%CITATION = CQGRD,27,075004;%%

%\cite{Loran:2009cr}
\bibitem{Loran:2009cr}
  F.~Loran and H.~Soltanpanahi,
  ``5D Extremal Rotating Black Holes and CFT duals,''
  Class.\ Quant.\ Grav.\  {\bf 26}, 155019 (2009)
  [arXiv:0901.1595 [hep-th]].
  %%CITATION = CQGRD,26,155019;%%

%\cite{Ghezelbash:2009gf}
\bibitem{Ghezelbash:2009gf}
  A.~M.~Ghezelbash,
  ``Kerr/CFT correspondence in the low energy limit of heterotic string theory,''
  JHEP {\bf 0908}, 045 (2009)
  [arXiv:0901.1670 [hep-th]].
  %%CITATION = JHEPA,0908,045;%%

%\cite{Lu:2009gj}
\bibitem{Lu:2009gj}
  H.~Lu, J.~w.~Mei, C.~N.~Pope and J.~F.~Vazquez-Poritz,
  ``Extremal static AdS black hole/CFT correspondence in gauged supergravities,''
  Phys.\ Lett.\  B {\bf 673}, 77 (2009)
  [arXiv:0901.1677 [hep-th]].
  %%CITATION = PHLTA,B673,77;%%

%\cite{Amsel:2009ev}
\bibitem{Amsel:2009ev}
  A.~J.~Amsel, G.~T.~Horowitz, D.~Marolf and M.~M.~Roberts,
  ``No Dynamics in the Extremal Kerr Throat,''
  JHEP {\bf 0909}, 044 (2009)
  [arXiv:0906.2376 [hep-th]].
  %%CITATION = JHEPA,0909,044;%%

%\cite{Compere:2009dp}
\bibitem{Compere:2009dp}
  G.~Compere, K.~Murata and T.~Nishioka,
  ``Central Charges in Extreme Black Hole/CFT Correspondence,''
  JHEP {\bf 0905}, 077 (2009)
  [arXiv:0902.1001 [hep-th]].
  %%CITATION = JHEPA,0905,077;%%

%\cite{Krishnan:2009tj}
\bibitem{Krishnan:2009tj}
  C.~Krishnan and S.~Kuperstein,
  ``A Comment on Kerr-CFT and Wald Entropy,''
  Phys.\ Lett.\  B {\bf 677}, 326 (2009)
  [arXiv:0903.2169 [hep-th]].
  %%CITATION = PHLTA,B677,326;%%

%\cite{Hotta:2009bm}
\bibitem{Hotta:2009bm}
  K.~Hotta,
  ``Holographic RG flow dual to attractor flow in extremal black holes,''
  Phys.\ Rev.\  D {\bf 79}, 104018 (2009)
  [arXiv:0902.3529 [hep-th]].
  %%CITATION = PHRVA,D79,104018;%%

%\cite{Astefanesei:2009sh}
\bibitem{Astefanesei:2009sh}
  D.~Astefanesei and Y.~K.~Srivastava,
  ``CFT Duals for Attractor Horizons,''
  Nucl.\ Phys.\  B {\bf 822}, 283 (2009)
  [arXiv:0902.4033 [hep-th]].
  %%CITATION = NUPHA,B822,283;%%

%\cite{Wen:2009qc}
\bibitem{Wen:2009qc}
  W.~Y.~Wen,
  ``Holographic descriptions of (near-)extremal black holes in five dimensional minimal supergravity,''
  arXiv:0903.4030 [hep-th].
  %%CITATION = ARXIV:0903.4030;%%

%\cite{Azeyanagi:2009wf}
\bibitem{Azeyanagi:2009wf}
  T.~Azeyanagi, G.~Compere, N.~Ogawa, Y.~Tachikawa and S.~Terashima,
  ``Higher-Derivative Corrections to the Asymptotic Virasoro Symmetry of 4d Extremal Black Holes,''
  Prog.\ Theor.\ Phys.\  {\bf 122}, 355 (2009)
  [arXiv:0903.4176 [hep-th]].
  %%CITATION = PTPKA,122,355;%%

%\cite{Wu:2009di}
\bibitem{Wu:2009di}
  X.~N.~Wu and Y.~Tian,
  ``Extremal Isolated Horizon/CFT Correspondence,''
  Phys.\ Rev.\  D {\bf 80}, 024014 (2009)
  [arXiv:0904.1554 [hep-th]].
  %%CITATION = PHRVA,D80,024014;%%

%\cite{Peng:2009wx}
\bibitem{Peng:2009wx}
  J.~J.~Peng and S.~Q.~Wu,
  ``Extremal Kerr/CFT correspondence of five-dimensional rotating (charged) black holes with squashed horizons,''
  Nucl.\ Phys.\  B {\bf 828}, 273 (2010)
  [arXiv:0911.5070 [hep-th]].
  %%CITATION = NUPHA,B828,273;%%


%%%% Asymptotic Symmetry  %%%%

%\cite{Brown:1986nw}
\bibitem{Brown:1986nw}
  J.~D.~Brown and M.~Henneaux,
  ``Central charges in the canonical realization of asymptotic symmetries: an example from three-dimensional gravity,''
  Commun.\ Math.\ Phys.\  {\bf 104}, 207 (1986).
  %%CITATION = CMPHA,104,207;%%


%%%%  Stress tensor  %%%%

%\cite{Hartman:2008dq}
\bibitem{Hartman:2008dq}
  T.~Hartman and A.~Strominger,
  ``Central charge for $AdS_2$ quantum gravity,''
  JHEP {\bf 0904}, 026 (2009)
  [arXiv:0803.3621 [hep-th]].
  %%CITATION = JHEPA,0904,026;%%

%\cite{Alishahiha:2008tv}
\bibitem{Alishahiha:2008tv}
  M.~Alishahiha and F.~Ardalan,
  ``Central charge for 2D gravity on AdS(2) and AdS(2)/CFT(1) correspondence,''
  JHEP {\bf 0808}, 079 (2008)
  [arXiv:0805.1861 [hep-th]].
  %%CITATION = JHEPA,0808,079;%%

%\cite{Castro:2008ms}
\bibitem{Castro:2008ms}
  A.~Castro, D.~Grumiller, F.~Larsen and R.~McNees,
  ``Holographic description of $AdS_2$ black holes,''
  JHEP {\bf 0811}, 052 (2008)
  [arXiv:0809.4264 [hep-th]].
  %%CITATION = JHEPA,0811,052;%%

%\cite{Balasubramanian:2009bg}
\bibitem{Balasubramanian:2009bg}
  V.~Balasubramanian, J.~de Boer, M.~M.~Sheikh-Jabbari and J.~Simon,
  ``What is a chiral 2d CFT? And what does it have to do with extremal black holes?,''
  JHEP {\bf 1002}, 017 (2010)
  [arXiv:0906.3272 [hep-th]].
  %%CITATION = JHEPA,1002,017;%%

%\cite{Castro:2010vi}
\bibitem{Castro:2010vi}
  A.~Castro, C.~Keeler and F.~Larsen,
  ``Three Dimensional Origin of $AdS_2$ Quantum Gravity,''
  arXiv:1004.0554 [hep-th].
  %%CITATION = ARXIV:1004.0554;%%


%%%%  Hidden Symmerty  %%%%%%%%%%%%%%%%%%%%%%%%%%%%%%%%%%%%%%%%%%%%%%%

%\cite{Castro:2010fd}
\bibitem{Castro:2010fd}
  A.~Castro, A.~Maloney and A.~Strominger,
  ``Hidden Conformal Symmetry of the Kerr Black Hole,''
  arXiv:1004.0996 [hep-th].
  %%CITATION = ARXIV:1004.0996;%%

%\cite{Krishnan:2010pv}
\bibitem{Krishnan:2010pv}
  C.~Krishnan,
  ``Hidden Conformal Symmetries of Five-Dimensional Black Holes,''
  arXiv:1004.3537 [hep-th].
  %%CITATION = ARXIV:1004.3537;%%


\end{references}
\end{document}